\documentstyle[12pt,epsfig]{article}
\def\be{\begin{equation}} \def\ee{\end{equation}} \def\bea{\begin{eqnarray}}
\def\eea{\end{eqnarray}} \def\nnb{\nonumber}

\begin{document}

\hfill{January 28, 2008}
%\hfill{January 28, 2008,\ \ \ {\tt ppfusion6}}

\begin{center}
\vskip 10mm 
{\Large\bf  Proton-proton fusion in pionless effective theory }
\vskip 10mm 

{\large 
S. Ando${}^{a,b}$\footnote{mailto:shung-ichi.ando@manchester.ac.uk}, 
J.~W. Shin${}^a$,
C.~H. Hyun$^c$, 
S.~W. Hong$^a$, 
and K. Kubodera$^d$ 
}
%Shung-ichi Ando\footnote{mailto:sando@meson.skku.ac.kr}, 
%Jae Won Shin %footnote{mailto:shine8199@hanmail.net},
%Chang Ho Hyun, %\footnote{mailto:hch@meson.skku.ac.kr},
%Seung Woo Hong, %\footnote{mailto:swhong@skku.ac.kr}
%Kuniharu Kubodera %\footnote{mailto:kubodera@physics.sc.edu}
%}%\\ 
\vskip 8mm 
{\large \it 
$^a$Department of Physics, 
Sungkyunkwan University,
Suwon 440-746, 
Korea 

$^b$Theoretical Physics Group,
School of Physics and Astronomy, \\
The University of Manchester,
Manchester, 
M13 9PL,
UK 

$^c$Department of Physics Education, 
Daegu University,
Gyeongsan 712-714, 
Korea

$^d$Department of Physics and Astronomy, 
University of South Carolina, 
Columbia, 
SC 29208, 
USA
}
\end{center}

\vskip 7mm

The proton-proton fusion reaction,
$pp\to de^+\nu$,
is studied
in pionless effective field theory (EFT)
with di-baryon fields up to next-to leading order.
With the aid of the di-baryon fields,
the effective range corrections are naturally resummed
up to the infinite order 
and thus the calculation is greatly simplified.  
Furthermore, the low-energy constant which appears in
the axial-current-di-baryon-di-baryon contact vertex
is fixed through the ratio of two- and one-body
matrix elements which reproduces the tritium lifetime
very precisely.
As a result we can perform a parameter free 
calculation for the process.
We compare our numerical result with those 
from the accurate potential model and previous pionless EFT calculations,  
and find a good agreement within the accuracy better than 1\%. 

\vskip 5mm
\noindent
{PACS(s): 21.45.Bc, 26.20.Cd, 26.65.+t}

\newpage 
\noindent {\bf 1. Introduction}

The proton-proton fusion process, $pp\to de^+\nu_e$,
is a fundamental reaction for the nuclear astrophysics,
especially important for the understanding of 
the star evolutions~\cite{b-pr38}
and solar neutrinos~\cite{b-89,aetal-rmp98,kp-arnps04}.
However, the process has never been studied experimentally 
because the event is extremely unlikely to take place
in the laboratory at the proton energies in the sun.
The calculation of the transition rate and its uncertainty
has naturally become a challenge to nuclear theory. 
The first calculation of the process was carried out 
by Bethe and Critchfield~\cite{bc-pr38} in 1938.
This estimation was improved by Salpeter~\cite{s-pr52}\footnote{
For a recent historical review, see Ref.~\cite{s-07}.} in 1952.
Later, small corrections, such as the electromagnetic radiative
corrections, were considered 
by Bahcall and his collaborators~\cite{bm-aj69,kb-aj94} 
in the framework of effective range theory.
Recently, accurate phenomenological potential models 
were employed to study the process~\cite{cetal-prc91,setal-prc98}. 
Furthermore, in Ref.~\cite{petal-aj98}
the two-nucleon current operators were calculated 
from heavy-baryon chiral perturbation theory (HB$\chi$PT)
up to next-to-next-to-next-to leading order 
(N$^3$LO), and Park {\it et al.} obtained
quite an accurate estimation ($\sim$ 0.3\% uncertaity) 
for the process
by fixing an unknown parameter,
so-called low energy constant (LEC), 
which appears in the two-nucleon-axial-current 
contact interaction 
in terms of the tritium lifetime~\cite{petal-03,petal-prc03}.

The kinetic energy relevant to the $pp$ fusion process 
at the core of the sun is quite low, 
$kT_c\simeq 1.18$ keV, where  $T_c$ is the core temperature
of the sun, $T_c\simeq 13.7\times 10^6$ K,
and $k$ is the Boltzmann constant. 
The proton momentum at the core, 
$p_c\simeq \sqrt{2m_pkT_c}\simeq 1.5$ MeV,
where $m_p$ is the proton mass,
is still significantly small 
compared to the pion mass,
$m_\pi \simeq 140$ MeV.
Therefore, we may regard the pion 
as a heavy degree of freedom for the $pp$ fusion process.
It may be convenient and suitable 
to employ a pionless effective field theory (EFT)~\cite{crs-npa99}, 
in which the pions are integrated out of the effective Lagrangian
for the process in question.
The $pp$ fusion process in the pionless theory 
has been studied by Kong and Ravndal~\cite{krs}
up to next-to leading order (NLO)
and by Butler and Chen~\cite{bc-plb01}
up to fifth order (N$^4$LO).
Thanks to the perturbative scheme in EFT,
the accuracy of the N$^4$LO calculation would,
in principle, be $(Q/\Lambda)^4\sim (1/3)^4\simeq 1$\%,
where $Q/\Lambda \sim 1/3$ is a typical expansion parameter
in the pionless theory. 
However, because of lack of the experimental data
to fix an unknown LEC $L_{1A}$
which appears in the two-nucleon-axial-current contact 
interaction in the pionless effective Lagrangian, 
an uncertainty estimated in the pionless EFT 
for the $pp$ fusion process is still significantly larger than
what is expected from the counting rules of the theory.

In this work, we employ a pionless EFT with 
di-baryon fields~\cite{bs-npa01,ah-prc05,st-07}.
\footnote{
We have employed the same formalism in the studies of
the two-body processes, such as neutron-neutron fusion~\cite{ak-plb06},
radiative neutron capture on a proton at BBN energies~\cite{aetal-prc06}, 
and neutral pion production in proton-proton collision 
near threshold~\cite{a-epja07}.}
The amplitude for the $pp$ fusion process at the 
zero proton momentum is calculated up to NLO.
We introduce two di-baryon fields~\cite{k-npb97}, 
which have the same quantum numbers
as those of $S$-wave two-nucleon states ($^1S_0$ and $^3S_1$ states),
as auxiliary fields: 
after integrating out the di-baryon fields 
we do have the ordinary pionless theory without the di-baryon fields.
However, 
as have intensively been discussed 
in Refs.~\cite{bs-npa01,ah-prc05,r0-resum,r0-resum2},
with the aid of the di-baryon fields, 
resummation of the effective range correction 
up to the infinite order is naturally introduced, 
which greatly simplifies the calculation of higher order
corrections to the wave functions.
In addition, the new counting rules make 
the expansion parameter $Q$ much improved,
and it is not necessary 
to employ the power divergence subtraction scheme~\cite{pds} 
any longer.    
Furthermore, by assuming
that the leading order (LO) contribution 
in the di-baryon-di-baryon-current contact interaction 
can be determined mainly from the one-body current interaction
as discussed in Ref.~\cite{ah-prc05}, 
we can reproduce 
the results from the effective range theory~\cite{ert}
in the LO calculations 
of the pionless EFT with the di-baryon fields.
The NLO correction, the di-baryon-di-baryon-current contact 
interaction denoted by the unknown LEC $l_{1A}$, 
is approximately presumed to be 
the two-body (2B) current correction in the pionful calculations. 
We fix the LEC $l_{1A}$ by using 
the relative strength of the two-body matrix element
to that of the one-body contribution,
$\delta_{2B}$~\cite{petal-prc03},
which has been determined from the accurate tritium lifetime datum.
(We discuss it in detail later.)
Consequently we can make our estimation of the $pp$ fusion amplitude
free from unknown parameters.
Moreover, though our calculation is rather simple and is only up to NLO,
we can obtain a result comparable
to that from the accurate potential model calculation
within the accuracy better than $\sim$ 1\%.

This paper is organized in the followings:
in Sec. 2, we introduce the pionless effective Lagrangian with
the di-baryon fields up to NLO, and
in Sec. 3, we fix the LECs which appear in the initial and final 
two-nucleon states by using the effective range parameters.
In Sec. 4, the amplitude for the $pp$ fusion process is calculated
up to NLO. We show our numerical results in Sec. 5. 
In Sec. 6, discussion and conclusions are given. 

\vskip 2mm \noindent
{\bf 2. Pionless effective Lagrangian with di-baryon fields}

For the low-energy process,
the weak-interaction Hamiltonian can be taken
to be
\bea
{\cal H} &=& \frac{G_FV_{ud}}{\sqrt{2}}l_\mu J^\mu\,  ,
\label{eq;H}
\eea
where $G_F$ is the Fermi constant and $V_{ud}$ is the
CKM matrix element.
$l_\mu$ is the lepton current
$l_\mu = \bar{u}_e \gamma_\mu (1-\gamma_5)v_\nu$,
and $J_\mu$ is the hadronic current.
We will calculate the two-body hadronic current $J^\mu$ 
from the pionless effective Lagrangian
with di-baryon fields 
up to NLO.

We adopt the standard counting rules
of pionless EFT with di-baryon fields~\cite{bs-npa01}.
Introducing an expansion scale $Q<\Lambda(\simeq m_\pi)$,
we count the magnitude of spatial part of the external and loop
momenta, $|\vec{p}|$ and $|\vec{l}|$, as $Q$,
and their time components, $p^0$ and $l^0$, as $Q^2$.
The nucleon and di-baryon propagators are of $Q^{-2}$,
and a loop integral carries $Q^5$.
The scattering lengths and effective ranges are counted as
$Q\sim \{\gamma, 1/a_0,1/\rho_d,1/r_0\}$
where $\gamma$, $a_0$, $\rho_d$ and $r_0$ are 
the effective range parameters for the $S$-wave $NN$ scattering;
$\gamma\equiv \sqrt{m_NB}$, where $B$ is the deuteron binding energy,
$a_0$ is the scattering length in the $^1S_0$ channel, 
$\rho_d$ and $r_0$ are the effective ranges in 
the $^3S_1$ and $^1S_0$ channel, respectively. 
The orders of vertices and transition amplitudes
are easily obtained by counting the numbers
of these factors in the Lagrangian
and diagrams, respectively.
As discussed below, some vertices acquire factors
like $r_0$ and $\rho_d$ after renormalization and thus
their orders can differ from what
the above naive dimensional analysis suggests.
Note that we do not include the higher order
radiative corrections,
such as the vacuum polarization effect~\cite{vp} 
and the radiative corrections from one-body part~\cite{aetal-plb04}. 

A pionless effective Lagrangian with di-baryon fields may
be written as~\cite{bs-npa01,ah-prc05}
\bea
{\cal L} &=& {\cal L}_N + {\cal L}_s + {\cal L}_t + {\cal L}_{st}\, ,
\eea
where ${\cal L}_N$ is a one-nucleon Lagrangian,
${\cal L}_s$ is the spin-singlet ($^1S_0$ state) di-baryon
Lagrangian including coupling to the two-nucleon,
${\cal L}_t$ is the spin-triplet ($^3S_1$ state) di-baryon
Lagrangian including coupling to the two-nucleon and
${\cal L}_{st}$ describes the weak-interaction transition
(due to the axial current) from the $^1S_0$ di-baryon to
the $^3S_1$ di-baryon.

A pionless one-nucleon Lagrangian
in the heavy-baryon formalism reads
\bea
{\cal L}_N &=&
N^\dagger \left\{ iv\cdot D -2ig_AS\cdot \Delta
+\frac{1}{2m_N}\left[(v\cdot D)^2 -D^2\right] +\cdots
\right\} N\, ,
\eea
where the ellipsis represents terms that do not appear
in this calculation.
$v^\mu$ is the velocity vector satisfying $v^2=1$;
we choose $v^\mu=(1,\vec{0})$, and
$S^\mu$ is the spin operator $2S^\mu = (0,\vec{\sigma})$.
Covariant derivative $D_\mu$ reads as
$D_\mu = \partial_\mu
-\frac{i}{2}\vec{\tau}\cdot \vec{\cal V}_\mu$
where $\vec{\cal V}_\mu$
is the external isovector vector current, and
$\Delta_\mu =
-\frac{i}{2}\vec{\tau}\cdot \vec{\cal A}_\mu$,
where $\vec{\cal A}_\mu$ is
the external isovector axial current.
$g_A$ is the axial-vector coupling constant
and $m_N$ is the nucleon mass.

The Lagrangians that involve the di-baryon fields
are given by
\bea
{\cal L}_s &=& \sigma_s s_a^\dagger \left[
iv\cdot D + \frac{1}{4m_N}[(v\cdot D)^2 -D^2] + \Delta_s \right] s_a
-y_s \left[ s_a^\dagger (N^TP_a^{(^1S_0)}N) +\mbox{\rm h.c.}
\right]\, ,
\\
{\cal L}_t &=& \sigma_t t_i^\dagger \left[
iv\cdot D + \frac{1}{4m_N}[(v\cdot D)^2 -D^2] + \Delta_t \right] t_i
-y_t \left[ t_i^\dagger (N^TP_i^{(^3S_1)}N) +\mbox{\rm h.c.}
\right]\, ,
\\
{\cal L}_{st} &=&
- \left[ \left(\frac{r_0+\rho_d}{2\sqrt{r_0\rho_d}}\right)\, g_A
+ \frac{l_{1A}}{m_N\sqrt{r_0\rho_d}}\right]
\left[ s_a^\dagger  t_i {\cal A}_i^a + \mbox{\rm h.c.}\right]\, ,
\label{eq;Lst}
\eea
where $s_a$ and $t_i$ are the di-baryon fields for the $^1S_0$ and
$^3S_1$ channel, respectively.
The covariant derivative for the di-baryon field is given by
$D_\mu = \partial_\mu -iC{\cal V}^{ext}_\mu$ where
${\cal V}_\mu^{ext}$ is the external vector field.
$C$ is the charge operator for the di-baryon field;
$C=0,1,2$ for the $nn$, $np$, $pp$ channel, respectively.
$\sigma_{s,t}$ is the sign factor 
$\sigma_{s,t}=\pm 1$
and $\Delta_{s,t}$ is the mass difference between the di-baryon
and two nucleons, $m_{s,t} = 2m_N+\Delta_{s,t}$.
$y_{s,t}$ is the di-baryon-two-nucleon coupling constant.
$P_i^{(S)}$ is the projection operator
for the $S$ = $^1S_0$ or $^3S_1$ channel;
\bea
P_a^{(^1S_0)} = \frac{1}{\sqrt8}\sigma_2\tau_2\tau_a\, ,
\ \ \
P_i^{(^3S_1)} = \frac{1}{\sqrt8}\sigma_2\sigma_i\tau_2\, ,
\ \ \
{\rm Tr}\left(P_i^{(S)\dagger}P_j^{(S)}\right)
= \frac12 \delta_{ij}\, ,
\eea
where $\sigma_i$ ($\tau_a$) is the spin (isospin) operator.
Note that, as mentioned in the Introduction, 
we separate the di-baryon-di-baryon-current contact interaction
in Eq.~(\ref{eq;Lst}) into
the LO and NLO terms.
The LO interaction proportional to $g_A$ is determined 
by the one-body axial-current interaction and the factor
$\frac12(r_0+\rho_d)/\sqrt{r_0\rho_d}$ is included 
so as to reproduce the result from the effective range theory
at LO.  
The NLO correction is parameterized by the LEC $l_{1A}$.
More detailed discussion about the separation of LO and NLO
contact interaction with external probe in the di-baryon formalism
can be found in Ref.~\cite{ah-prc05}.

\vskip 2mm \noindent
{\bf 3. Initial and final $NN$ channels }

The typical energy of the $pp$ fusion reaction 
is very low, as discussed in the Introduction, 
so we can assume that the dominant channel
of the reaction is from the initial $^1S_0$ $pp$ state
to the final $^3S_1$ deuteron state. 
In this section, we fix the LECs which appear in 
the initial and final two-nucleon states for 
the $pp$ fusion process from the effective range parameters.

\begin{figure}
\begin{center}
\epsfig{file=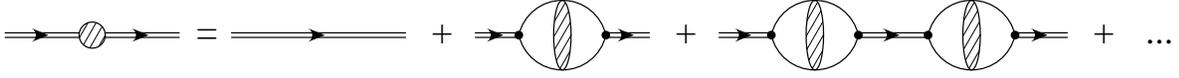,width=15.5cm}
\caption{
Diagrams for the dressed di-baryon propagator
including the Coulomb interaction.
A double-line with a filled circle denotes
the renormalized dressed di-baryon propagator.
Double-lines without the filled circle and
single-curves
denote the bare di-baryon propagators and
nucleon propagators, respectively.
Two-nucleon propagator with
a shaded blob denotes the Green's function
including the Coulomb potential.
A (spin-singlet) di-baryon-nucleon-nucleon ($sNN$) vertex is proportional
to the LEC $y_s$.
\label{fig;dibaryon-propagator}}
\end{center}
\end{figure}

In Fig.~\ref{fig;dibaryon-propagator},
LO diagrams for the initial $pp$ state 
in $^1S_0$ channel, i.e.,
the dressed $^1S_0$ channel di-baryon propagator,
are shown where the two-nucleon bubble diagrams
including the Coulomb interaction
are summed up to
the infinite order.
The inverse of the propagator
in the center of mass (CM) frame is thus obtained by
\bea
i D_{s}^{-1}(p) &=&
i\sigma_{s}(E+\delta_{s})-iy_{s}^2J_0(p)  \, ,
\eea
with
\bea
J_0(p) &=&
\int\frac{d^3\vec{k}}{(2\pi)^3}
\frac{d^3\vec{q}}{(2\pi)^3}
\langle \vec{q}| \hat{G}_C^{(+)}(E)|\vec{k}\rangle\, ,
\eea
where $\hat{G}_C^{(+)}$ is the outgoing two-nucleon
Green's function including the Coulomb potential,
\bea
\hat{G}_C^{(+)}(E)=\frac{1}{E-\hat{H}_0-\hat{V}_C+i\epsilon}\, .
\eea
$E$ is the total CM energy, $E\simeq p^2/m_N$, 
$\hat{H}_0$ is the free Hamiltonian for two-proton,
$\hat{H}_0=\hat{p}^2/m_N$, and $\hat{V}_C$ is the repulsive
Coulomb force $\hat{V}_C= \alpha/r$: $\alpha$ is the
fine structure constant.
Employing the dimensional regularization
in $d=4-2\epsilon$ space-time dimension,
we obtain~\cite{kr-plb99,ashh-07}
\bea
J_0(p) &=&
\frac{\alpha m_N^2}{8\pi}\left[
\frac{1}{\epsilon}
-3C_E 
+ 2
+{\rm ln}\left(\frac{\pi\mu^2}{\alpha^2m_N^2}\right)
\right]
-\frac{\alpha m_N^2}{4\pi}h(\eta)
-C_\eta^2\frac{m_N}{4\pi}(ip)\, ,
\eea
where $\mu$ is the scale of the dimensional regularization,
$C_E= 0.5772\cdots$,
and
\bea
&& h(\eta) = Re\, \psi(i\eta)-{\rm ln}\eta\, ,
\ \ \
Re\, \psi(\eta) =
\eta^2 \sum_{\nu=1}^\infty \frac{1}{\nu(\nu^2+\eta^2)}
-C_E \, ,
\nnb \\ &&
C_\eta^2 = \frac{2\pi \eta}{e^{2\pi\eta}-1}\, ,
\ \ \
\eta= \frac{\alpha m_N}{2p}\, .
\eea
Thus the inverse of renormalized dressed di-baryon propagator
is obtained as
\bea
i D_{s}^{-1}(p) &=&
iy_s^2 \frac{m_N}{4\pi}\left[
\frac{4\pi \sigma_s \Delta_s^R}{m_Ny_s^2}
+ \frac{4\pi\sigma_s}{m_N^2y_s^2}p^{2}
+\alpha m_N h(\eta)
+ ip\, C_\eta^2
\right]\, ,
\eea
where $\Delta_s^R$ is the renormalized
mass difference
\bea
\sigma_s \Delta_s^R &=&
\sigma_s \Delta_s
-y_s^2
\frac{\alpha m_N^2}{8\pi}\left[
\frac{1}{\epsilon}
-3C_E 
+ 2
+{\rm ln}\left(\frac{\pi\mu^2}{\alpha^2m_N^2}\right)
\right] \, .
\eea

\begin{figure}[t]
\begin{center}
\epsfig{file=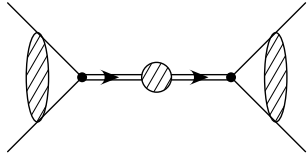,width=4cm}
\caption{
Diagram for the $S$-wave $pp$ scattering amplitude
with the Coulomb and strong interactions.
See the caption of Fig.~\ref{fig;dibaryon-propagator}
for details.
\label{fig;NNamplitudes}}
\end{center}
\end{figure}
In Fig.~\ref{fig;NNamplitudes},
a diagram of the $S$-wave $pp$ scattering
amplitude with the Coulomb and strong interactions is shown.
Thus we have the $S$-wave scattering amplitude as
\bea
i A_{s} &=&
(-iy_{s}\psi_0)iD_{s}(p)(-iy_{s}\psi_0)
\nnb \\
&=&
i \frac{4\pi}{m_N}\frac{C_\eta^2e^{2i\sigma_0}}{
-\frac{4\pi\sigma_{s}\Delta_{s}^R}{m_Ny_{s}^2}
-\frac{4\pi\sigma_{s} p^2}{m_N^2y_{s}^2}
-\alpha m_N h(\eta)
-ip\, C_\eta^2} \, ,
\eea
with
\bea
\psi_0 &=&
\int\frac{d^3\vec{k}}{(2\pi)^3}
\langle \vec{k}|\psi_{\vec{p}}^{(+)}\rangle
=\int\frac{d^3\vec{k}}{(2\pi)^3}
\langle\psi_{\vec{p}}^{(-)}|\vec{k}\rangle
= C_\eta e^{i\sigma_0}\, ,
\eea
where $\langle\vec{k}|\psi_{\vec{p}}^{(\pm)}\rangle$ are
the Coulomb wave functions
obtained by solving the Schr\"{o}dinger equations
$(\hat{H}-E)|\psi^{(\pm)}_{\vec{p}}\rangle = 0$
with $\hat{H}=\hat{H}_0+\hat{V}_C$
and represented in the $|\vec{k}\rangle$ space
for the two protons.
$\sigma_0$ is the $S$-wave Coulomb phase shift
$\sigma_0={\rm arg}\,\Gamma(1+i\eta)$.
The $S$-wave amplitude $A_s$ is
given in terms of the effective range parameters
as
\bea
i A_s &=&
i \frac{4\pi}{m_N}\frac{C_\eta^2e^{2i\sigma_0}}{
-\frac{1}{a_C} + \frac12 r_0 p^2 + \cdots
-\alpha m_N h(\eta)
-ip\, C_\eta^2} \, ,
\eea
where $a_C$ is the scattering length,
$r_0$ is the effective range,
and the ellipsis represents
the higher order effective range corrections.
Now it is easy to match the parameters
$\sigma_s$ and $y_s$ with the effective range parameters.
Thus we have $\sigma_{s} = -1$ and
\bea
&& y_s = \pm \frac{2}{m_N}\sqrt{\frac{2\pi}{r_0}}\, ,
\ \ \
D_s(p) = \frac{m_Nr_0}{2}\frac{1}{\frac{1}{a_C}
-\frac12r_0 p^2 +\alpha m_N h(\eta) +ip\, C_\eta^2}\, .
\label{eq;ys}
\eea

\begin{figure}
\begin{center}
\epsfig{file=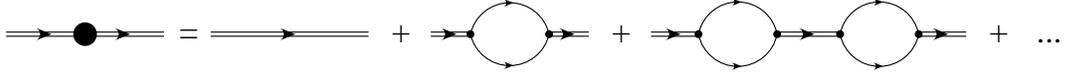,width=14cm}
\caption{
Dressed di-baryon propagator without Coulomb
interaction (double line with a filled circle)
at leading order.
A single line stands for the nucleon,
while a double line represents the bare di-baryon.}
\label{fig;NN-d-prop}
\end{center}
\end{figure}
In Fig.~\ref{fig;NN-d-prop},
LO diagrams for the final deuteron channel, i.e.,
the dressed $^3S_1$ channel di-baryon propagators
are depicted. 
Since insertion of a two-nucleon one-loop diagram 
does not alter the order of the diagram, 
the two-nucleon bubbles
should be summed up to the infinite order.
Thus the inverse of the dressed di-baryon propagator
for the deuteron channel in the CM frame reads
\bea
iD_{t}^{-1}(p) &=& i\sigma_{t} (E+\Delta_{t})
+ i y_{t}^2 \frac{m_N}{4\pi} (ip)
\nnb \\ &=&
i\frac{m_Ny_{t}^2}{4\pi}
\left[
  \frac{4\pi\sigma_{t} \Delta_{t}}{m_Ny^2_{t}}
+ \frac{4\pi\sigma_{t}E}{m_Ny_{t}^2}+ip
\right]\, ,
\eea
where we have used dimensional regularization for
the loop integral and $E$ is the total energy of the
two nucleons,
$E\simeq p^2/m_N$.
\begin{figure}
\begin{center}
\epsfig{file=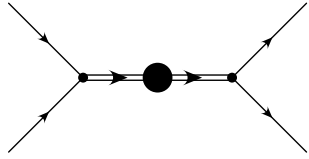,width=4cm}
\caption{
Diagram for the $S$-wave $NN$ amplitude without Coulomb interaction
at leading order.
The double line with a filled circle represents the
dressed di-baryon
propagator obtained in Fig.~\ref{fig;NN-d-prop}.
}
\label{fig;NN}
\end{center}
\end{figure}
The dressed di-baryon propagators 
are renormalized via the $S$-wave $NN$ amplitudes.
The amplitudes obtained from the diagram
in Fig.~\ref{fig;NN}
should satisfy
\bea
i A_{t} &=& (-iy_{t})\left[
iD_{t}(p)\right] (-iy_{t})
= \frac{4\pi}{m_N}
  \frac{i}{-\frac{4\pi \sigma_{t}\Delta_{t}}{m_Ny_{t}^2}
- \frac{4\pi \sigma_{t}}{m_Ny_{t}^2} p^2
- ip } \, ,
\eea
where $A_{t}$ is related to the $S$-wave $NN$ scattering
$S$-matrix via
\bea
S-1 = e^{2i\delta_{t}}-1 = \frac{2ip}{p\, {\rm cot}\delta_{t}-ip}
= i\left(\frac{pm_N}{2\pi}\right) A_{t}\, .
\eea
Here $\delta_t$ is the phase shift
for the $^3S_1$ channel.
Meanwhile, effective range expansion reads
\bea
p\, {\rm cot}\delta_t = - \gamma + \frac12 \rho_d(\gamma^2+p^2) + \cdots\,.
\eea
%where %$\gamma$ is the deuteron momentum
%$\gamma\equiv \sqrt{m_NB}$ ($B$ is the deuteron binding energy),
%and $\rho_d$ is the effective range for the bound $^3S_1$ channel.
%
Now, the above renormalization condition
allows us to relate the LECs
to the effective-range expansion parameters.
For the deuteron channel, one has
$\sigma_t=-1$ and
\bea
y_t = \pm \frac{2}{m_N}\sqrt{\frac{2\pi}{\rho_d}}\, ,
\ \ \
D_t(p) = \frac{m_N\rho_d}{2}
\frac{1}{\gamma+ip-\frac12 \rho_d(\gamma^2+p^2)}
= \frac{Z_d}{E+B} +\cdots \, ,
\label{eq;Dt}
\eea
where $Z_d$ is the wave function normalization factor
of the deuteron at the pole $E=-B$,
and the ellipsis in Eq.~(\ref{eq;Dt})
denotes corrections that are finite or vanish at
$E=-B$.
Thus one has \cite{bs-npa01}
\bea
Z_d = \frac{\gamma\rho_d}{1-\gamma\rho_d}\, .
\eea
This $Z_d$ is equal to
the asymptotic $S$-state normalization constant.
It is to be noted
that the order of the
LECs $y_{t}$ is now of $Q^{1/2}$, and
the deuteron state is also described
by the renormalized dressed di-baryon propagator.

\vskip 2mm \noindent
{\bf 4. Amplitude up to NLO}

\begin{figure}[t]
\begin{center}
\epsfig{file=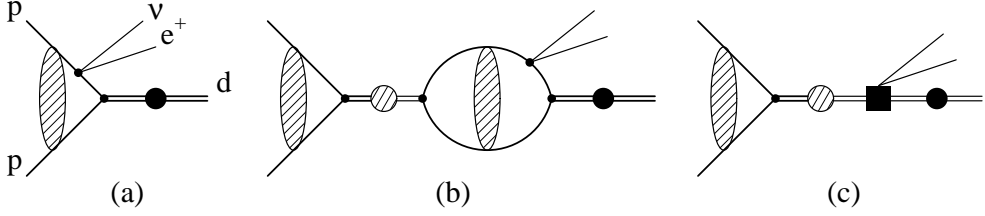,width=13cm}
\caption{
Diagrams for the $pp$ fusion process,
$pp\to de^+\nu_e$, up to NLO. }
\label{fig;ppfusion}
\end{center}
\end{figure}
Diagrams for the $pp$ fusion process up to NLO 
are shown in Fig.~\ref{fig;ppfusion}.
In the limit $p\to 0$, we have the amplitude from the
diagrams in the figure as
\bea
A &=& - \vec{\epsilon}_{(d)}^*\cdot \vec{\epsilon}_{(l)}
G_FV_{ud}\, g_A \, T_{fi}\, .
\eea
Here $\vec{\epsilon}_{(d)}^*$ is the spin polarization vector
of the out-going deuteron, $\vec{\epsilon}_{(l)}$ is the spatial part
of the lepton current $l^\mu$ in Eq.~(\ref{eq;H}), and 
\bea
T_{fi} &\simeq& \sqrt{\frac{8\pi\gamma}{1-\gamma\rho_d}}
\frac{C_\eta e^{i\sigma_0}}{\gamma^2}  
\left[
e^\chi -a_C\gamma\chi I(\chi)
+ \frac14 a_C (r_0+\rho_d)\gamma^2
+ \frac{a_C\gamma^2}{2g_Am_p}l_{1A}
\right]\, ,
\eea
where 
\bea
I(\chi)  
= \frac{1}{\chi}-e^\chi E_1(\chi) \, ,
\ \ \ 
\ \ \ 
E_1(\chi) = \int^\infty_\chi dt\, \frac{e^{-t}}{t}\, ,
\eea
with $\chi=\alpha m_p/\gamma$. 
We note that the amplitude $T_{fi}$ vanishes at the $p\to 0$ limit
because of the overall factor $C_\eta$.
The approximation is taken by keeping $p$ dependence in $C_\eta$
while ignoring higher order $p/m_N$ corrections in the remaining
part. Since $p/m_N \sim 0.2$ \%, the contribution from the higher
order $p/m_N$ terms will be sub 1 \% order, which can be neglected
conservatively at the uncertainty level we are considering in the 
present work.
Introducing a ``standard reduced matrix element''~\cite{krs},
\bea
\Lambda(p) = \sqrt{\frac{\gamma^3}{8\pi C_\eta^2}}\, |T_{fi}(p)|\, ,
\eea
we have a finite and analytic expression 
of the reduced matrix element $\Lambda(p)$ 
in the $p\to 0$ limit 
as
\bea
\Lambda(0) = \frac{1}{\sqrt{1-\gamma\rho_d}}
\left\{
e^\chi -a_C\gamma 
\left[1-\chi e^\chi E_1(\chi)\right]
+ \frac14 a_C (r_0+\rho_d)\gamma^2
+ \frac{a_C\gamma^2}{2g_Am_p}l_{1A}
\right\}\, .
\label{eq;Lambda0}
\eea
As mentioned above, 
we exactly reproduce the result of the effective range theory
at LO, and have a higher order correction 
proportional to the LEC $l_{1A}$ at NLO
in Eq.~(\ref{eq;Lambda0}).

\vskip 2mm \noindent
{\bf 5. Numerical results}

We obtain the matrix element $\Lambda(0)$ in Eq.~(\ref{eq;Lambda0})
in terms of the four effective range parameters, 
$a_C$, $r_0$, $\gamma$ and $\rho_d$, and the LEC $l_{1A}$.
The values of the effective range parameters are well known, 
but three of them are slightly different in the references.
In this work, we take two sets of the values: one is 
$a_C=-7.8063\pm 0.0026$ fm,
$r_0= 2.794\pm 0.014$ fm, and
$\rho_d=1.760\pm 0.005$ fm 
from Table VIII in Ref.~\cite{av18}. 
The other is
$a_C=-7.8149\pm 0.0029$ fm,
$r_0= 2.769\pm 0.014$ fm, and
$\rho_d=1.753\pm 0.008$ fm 
from Table XIV in Ref.~\cite{cdborn}. 
We take an average of numerical values of $\Lambda(0)$ 
from the two sets of the parameters for our numerical result.
The value of the LEC $l_{1A}$ should be fixed by 
experimental data, but there are no precise ones for
the two-body system.
We fix the value of the LEC $l_{1A}$
indirectly from the relative strength of the two-body matrix 
element to one-body one, 
$\delta_{2B}\equiv {\cal M}_{2B}/{\cal M}_{1B} =
(0.86\pm 0.05)$ \% in Eq.~(29) in Ref.~\cite{petal-prc03}.
This value has been obtained 
from the accurate potential model calculation
for the two-body matrix element
with the current operators derived from HB$\chi$PT up to N$^3$LO
where the two-body current operator has been fixed 
from an accurate experimental datum,
the tritium lifetime, for the three-body system.
Thus we have 
\bea
l_{1A} = -0.50 \pm 0.03\, ,
\eea
where we have used our LO amplitude as the one-body input.
This is a good approximation because the difference between
the amplitude from the effective range theory, which is almost 
the same as our LO result, and that from accurate potential model 
calculations is tiny~\cite{petal-aj98}.
For other well known parameters, we use $B=2.224575$ MeV, $g_A=1.2695$,
$m_p=938.272$ MeV, and $m_n=939.565$ MeV, and thus have
$\gamma= 45.70$ MeV, $\chi= 0.1498$, and $E_1(\chi)=1.465$.

Employing the values of the parameters mentioned above,
we have $\Lambda_{\rm LO}(0)=2.641$ at LO, and 
$\Lambda_{\rm NLO1}(0)=2.662\pm 0.002$ from the first set of 
the parameter values  
and $\Lambda_{\rm NLO2}(0)=2.664\pm 0.003$ from the second one
up to NLO. Thus we have an average value 
\bea
\Lambda_{\rm NLO}(0) = 2.663 \pm 0.004\, ,
\eea
and $\Lambda^2_{\rm NLO}(0) = 7.09\pm 0.02$ where
the estimated error bars mainly come from those 
of the effective ranges, $r_0$ and $\rho_d$, and the LEC $l_{1A}$.

\begin{table}
\begin{center}
\begin{tabular}{c || c  c c c} \hline
              & Our result & KR(NLO)\cite{krs} 
     & BC(N$^4$LO)\cite{bc-plb01} & Pot. model\cite{setal-prc98} \\ \hline
$\Lambda^2(0)$&7.09$\pm$0.02 & 7.04$\sim$7.70 
     & 6.71$\sim$7.03 & 7.05$\sim$7.06  \\ \hline
\end{tabular}
\end{center}
\caption{\label{tab;Lambda02}
Estimated values of $\Lambda^2(0)$.
The value in second column is our result.
The values in third, fourth, and fifth column
are estimated from the pionless EFT calculation up to NLO
by Kong and Ravndal (KR)~\cite{krs},
that up to N$^4$LO by
Butler and Chen (BC)~\cite{bc-plb01},
and an accurate phenomenological potential model 
calculation~\cite{setal-prc98}, respectively. }
\end{table}
In Table \ref{tab;Lambda02}, we compare our
numerical result for $\Lambda^2(0)$ with those from
other theoretical estimations,
the pionless EFT without di-baryons up to NLO
by Kong and Ravndal (KR)~\cite{krs},
that up to N$^4$LO by Butler and Chen (BC)~\cite{bc-plb01},
and the accurate phenomenological potential model 
calculation~\cite{setal-prc98}. 
We find that our numerical result is in good agreement with
the values from the former theoretical estimations
within the accuracy less than 1 \%.
As discussed before, the uncertainties of the estimations
from the pionless EFT without di-baryon fields are
still large, $\sim$4.5 \% for the KR's estimation up to NLO,
and $\sim$2.3 \% for the BC's one up to N$^4$LO,
mainly because of the unfixed LEC $L_{1A}$. 
Though the results in the previous pionless EFT calculations
have the unfixed LEC $L_{1A}$, 
we can directly compare our result of 
the amplitude $\Lambda(0)$ in Eq.~(\ref{eq;Lambda0}) to 
the expressions in Eq. (7) in Ref.~\cite{bc-plb01},
and fix the value of the LEC $L_{1A}$. 
Assuming the higher order LEC $\overline{K}_{1A}=0$,
we have $L_{1A}= 1.27\pm 0.12$ fm$^3$, which is consistent with
our previous estimation, $L_{1A}=1.18\pm 0.11$ fm$^3$ in 
Ref.~\cite{ak-plb06}. 
When comparing our result with that from the accurate phenomenological
potential model calculation, we find that our result 
is overestimated by $\sim$0.5 \%
mainly because we have not included the important contribution
from the vacuum polarization effect.

As a last remark we would like to note that the precedent 
pionless EFT calculations
include the higher order corrections in both wave functions and
vertices with external probe.
The contribution to $\Lambda(0)$ from the wave functions read
2.51, 2.54 and 2.58 at LO, NLO and N$^4$LO, respectively.
In our calculation with di-baryon field, higher order corrections
to the wave functions are incorporated naturally by the summation of
effective range contribution to infinite order, which gives
$\Lambda(0)$ equal to 2.64.
A great advantage of the pionless EFT with di-baryon field lies in
that we don't need to care the higher order contribution to
the wave function, and it is sufficient to take into account only 
the corrections to the vertices with external probe.
This advantage reduces the number of Feynman diagrams dramatically,
and makes the calculation of higher order terms very simple.

\vskip 2mm \noindent
{\bf 6. Discussion and conclusions}

In this work, we employed the pionless EFT with di-baryon fields
including the Coulomb interaction, and calculated the analytic 
expression of the amplitude
for the $pp$ fusion process, $pp\to de^+\nu_e$, up to NLO.
Employing the assumption to distinguish LO and NLO terms
in the contact di-baryon-di-baryon-axial-current interaction,
we reproduced the expression for the amplitude 
of the effective range theory at LO.
The LEC $l_{1A}$, which appears in the contact 
di-baryon-di-baryon-axial-current
interaction at NLO, is fixed by using the relative strength of
the two-body amplitude to the one-body one, $\delta_{2B}$, which
has been determined from the tritium lifetime 
in the HB$\chi$PT calculation,
and thus we could perform the parameter-free-calculation 
for the $pp$ fusion process.
We find that our numerical result of squared reduced amplitude
$\Lambda^2(0)$ is in good agreement with those of the 
recent theoretical calculations within the accuracy 
better than 1\%. 

As mentioned in the Introduction,
the current theoretical uncertainties for the $pp$ fusion 
process is $\sim$ 0.3\% in the HB$\chi$PT calculaiton
up to N$^3$LO~\cite{petal-prc03}. 
To improve our result to a few tenth \% accuracy, 
it would be essential to include the higher order corrections
in the modified counting rules discussed in the neutron 
beta decay calculation~\cite{aetal-plb04}:
the next higher order corrections would be 
the $\alpha$ order and $1/m_N$ corrections.
It is known that the higher $\alpha$ order corrections, 
such as the vacuum polarization effect~\cite{kb-aj94} 
and the radiative corrections from 
the one-body part~\cite{aetal-plb04}\footnote{
The radiative corrections from the one-body part
are quite significant, $2\sim 3$\% level,
and are conventionally included into the renormalized Fermi constant
$G_V'\simeq G_FV_{ud}$ and the phase factor $f_{pp}$
in the estimation of the $S$ factor for the $pp$ fusion process.
}, 
are significant, 
whereas the corrections from the $1/m_N$ terms 
would be $p_c/m_N\sim 0.16$\%. 
It would be worth calculating the $S$ factor for the $pp$ fusion
process in a few tenth \% accuracy
with the pionless EFT with di-baryon fields
including those higher order corrections. 

Another issue that we would need to clarify
is the value of the LEC $l_{1A}$, which has been fixed 
in this work by
using the result from the HB$\chi$PT calculation. 
As discussed, e.g., in Refs.~\cite{petal-prc03,gp-prl06},
the LECs which appear in the two-di-baryon-axial-current 
or four-nucleon-axial-current contact interactions,
denoted by $l_{1A}$ in the pionless EFT with di-baryon fields,
$L_{1A}$ in the pionless EFT without di-baryon fields,
and $\hat{d}^R$ in HB$\chi$PT,
are universal.
In other words, those LECs are shared 
by the processes, such as, 
the $pp$ fusion process 
($pp \to de^+\nu_e$)~\cite{petal-aj98,petal-03,petal-prc03,krs,bc-plb01},  
$nn$ fusion process ($nn\to de^-\bar{\nu}_e$)~\cite{ak-plb06},
neutrino deuteron reactions 
($\nu_e d\to pp e^-$, $\nu_e d\to np\nu_e$)~\cite{aetal-plb03,bc-npa00},
muon capture on the deuteron 
($\mu^- d \to nn \nu_\mu$)~\cite{aetal-plb02,cetal-prc05},
radiative pion capture on the deuteron ($\pi^-d\to nn\gamma$~\cite{gp-prc06} 
and its crossed partner $\gamma d\to nn\pi^+$~\cite{letal-epja05}),
tritium beta decay~\cite{petal-prc03}, and 
hep process ($p\,\mbox{$^3$He}\to \mbox{$^4$He}\,e^+\nu_e$)~\cite{petal-prc03}. 
If these LECs are determined by using the experimental data
from one of the processes, the lattice simulation~\cite{ds-npa04}, 
or the renormalization group method~\cite{na-prc06}, 
then we can predict the other processes 
in each of the formalisms without any unknown parameters. 
In this respect, it may be worth fixing the LEC $l_{1A}$  
in the same formalism, the pionless EFT with di-baryon fields,
from, e.g., the tritium lifetime 
extending our formalism to the three-body systems
with electroweak external probes.

\vskip 2mm \noindent
{\bf Acknowledgments}

We would like to thank T.-S. Park for communications
and M.~C. Birse for reading the manuscript and 
commenting on it.
SA is supported by the Korean Research Foundation and
the Korean Federation of Science and Technology Societies Grant
funded by Korean Government (MOEHRD, Basic Research Promotion Fund):
the Brain Pool program (052-1-6) and KRF-2006-311-C00271,
and by STFC grant number PP/F000448/1.
Work of JWS and SWH is supported by the Korea Science and Engineering
Foundation grant funded by the Korean Government (MOST) 
(No. M20608520001-07B0852-00110).
The work of KK is supported by the US National Science Foundation 
under Grant PHY-0457014.

\end{document}